\documentclass{article}

\usepackage[dblblindworkshop, final]{neurips_2025}
\workshoptitle{Foundation Models for the Brain and Body}

\usepackage{graphicx}
\usepackage[utf8]{inputenc} % allow utf-8 input
\usepackage[T1]{fontenc}    % use 8-bit T1 fonts
\usepackage{hyperref}       % hyperlinks
\usepackage{url}            % simple URL typesetting
\usepackage{booktabs}       % professional-quality tables
\usepackage{amsfonts}       % blackboard math symbols
\usepackage{nicefrac}       % compact symbols for 1/2, etc.
\usepackage{microtype}      % microtypography
\usepackage{xcolor}         % colors

\title{Mouse-Guided Gaze: Semi-Supervised Learning of Intention-Aware Representations for Reading Detection}

\author{%
  Seongsil Heo\\
  Department of Computer Science and Engineering\\
  University of California, Santa Cruz\\
  Santa Cruz, CA 95064 \\
  \texttt{sheo1@ucsc.edu} \\
  \And
  Roberto Manduchi \\
  Department of Computer Science and Engineering\\
  University of California, Santa Cruz\\
  Santa Cruz, CA 95064 \\
  \texttt{manduchi@soe.ucsc.edu} \\
}

\begin{document}

\maketitle

\begin{abstract}
Understanding user intent during magnified reading is critical for accessible interface design. Yet magnification collapses visual context and forces continual viewport dragging, producing fragmented, noisy gaze and obscuring reading intent. We present a semi-supervised framework that learns intention-aware gaze representations by leveraging mouse trajectories as weak supervision. The model is first pretrained to predict mouse velocity from unlabeled gaze, then fine-tuned to classify reading versus scanning. To address magnification-induced distortions, we jointly model raw gaze within the magnified viewport and a compensated view remapped to the original screen, which restores spatial continuity across lines and paragraphs. Across text and webpage datasets, our approach consistently outperforms supervised baselines, with semi-supervised pretraining yielding up to 7.5\% F1 improvement in challenging settings. These findings highlight the value of behavior-driven pretraining for robust, gaze-only interaction, paving the way for adaptive, hands-free accessibility tools.
\end{abstract}

\section{Introduction}

Modeling human behavior from multimodal signals is a central goal in human-computer interaction. Among such signals, eye movements play a critical role in understanding how users engage with text \cite{kaakinen2018fluctuation, miller2015using, raney2014using}. In screen magnification environments, the visible area is restricted to only a few words or lines at a time~\cite{blenkhorn2003screen}. To follow the text, users continuously drag the viewport with the mouse, moving horizontally to keep the current line in view and vertically to reach the next line. As a result, the mouse becomes indispensable yet taxing, especially for people with low vision who rely on screen magnifiers~\cite{lee2021bringing, lee2012gesture}. This burden motivates automatic scroll control to reduce effort and preserve reading flow. 

Intuitive and precise automatic scroll control requires accurate, low-latency inference of user intent. A key step is distinguishing focused reading from exploratory scanning. This distinction helps systems determine when and how to scroll based on the user's reading state, enabling more structured control policies that adapt to behavioral context~\cite{turner2015understanding}. Although several gaze-based automatic scroll control systems have been proposed, most rely on handcrafted heuristics that generalize poorly across users and content types~\cite{manduchi2022gaze, sharmin2013reading}. To address these limitations, we propose a behavior classification model that robustly separates reading from scanning using gaze signals.

\begin{figure}[t]
  \centering
  \includegraphics[width=\textwidth]{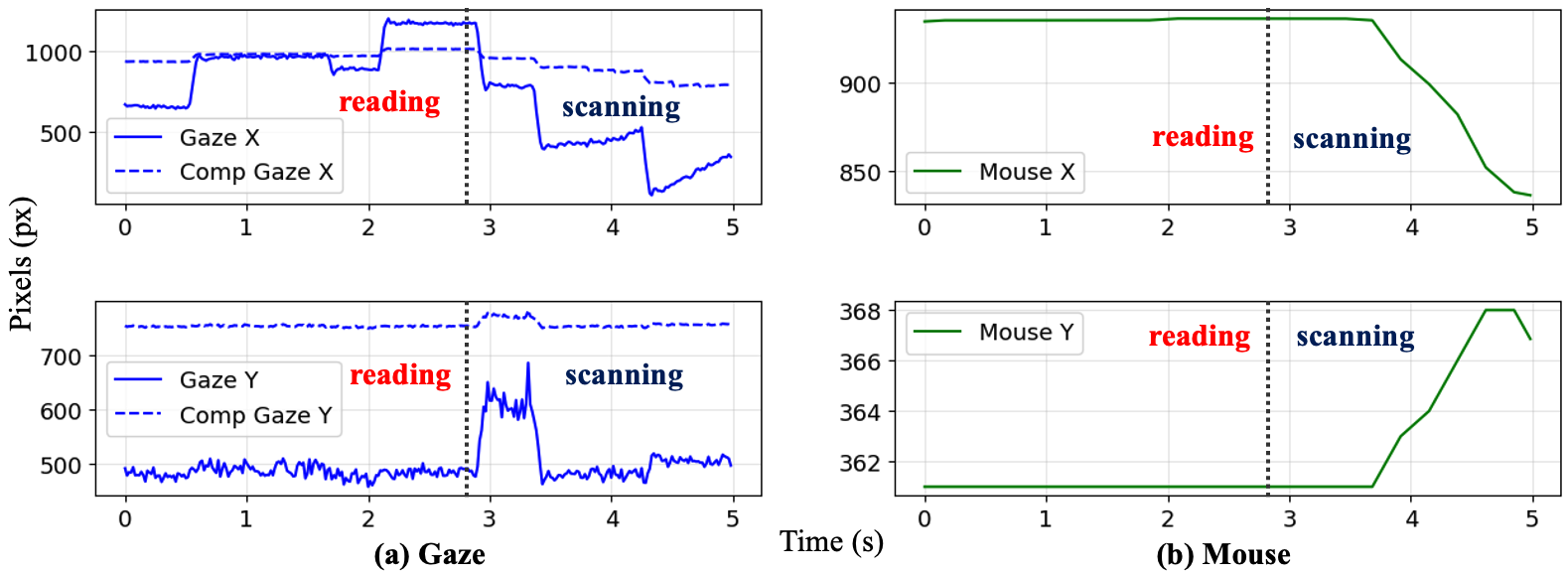}
  \vspace*{-0.5em}
  \caption{Five second segment from our dataset that spans a transition from reading to scanning (marked by a vertical dashed line). (a) Gaze position. Top panel shows X, bottom panel shows Y. Raw gaze is plotted with a solid line and the compensated gaze with a dashed line; “Comp” denotes compensated. (b) Mouse cursor position. Top panel shows X, bottom panel shows Y, with motion increases during scanning.}
  \label{fig:data_vis}
  \vspace{-6pt}
\end{figure}

Distinguishing between reading and scanning from gaze alone is challenging, especially under screen magnification. In such settings, gaze trajectories are often fragmented and noisy due to limited visual context, viewport shifts, and disrupted eye movement patterns~\cite{heo2024reading, manduchi2022gaze, maus2020gaze}. Fig.~\ref{fig:data_vis} (a) illustrates an example segment of gaze and compensated gaze. Gaze captures local eye movements relative to the magnified viewport, while compensated gaze remaps these coordinates back to the original screen space using the magnification factor\cite{heo2024reading, manduchi2024eye}. As shown, raw gaze appears irregular and noisy, while the compensated trace forms smoother, line-aligned trajectories, underscoring the difficulty of the task and the need for robust modeling.

To improve robustness, we use two complementary views of gaze. We jointly model raw and compensated gaze for behavior classification, capturing fine-grained movements while maintaining global spatial consistency in magnified settings. To our knowledge, this is the first approach to fuse these two streams for this task.

In addition, we introduce a semi-supervised learning framework that leverages mouse input only during training to guide the learning of gaze representations. As shown in Fig.~\ref{fig:data_vis} (b), mouse activity tends to increase during scanning episodes, reflecting users’ need to reorient or reposition the viewport. Because mouse trajectories reflect deliberate, task-driven actions~\cite{huang2012improving}, they provide informative supervision even when gaze is noisy. Concretely, we pretrain the model to predict mouse velocity from unlabeled gaze and then fine-tune it for intent classification on a labeled set. At inference, it operates solely on gaze, enabling real-time, hands-free classification for accessibility.

\section{Related Work}

%This remapping restores the global layout and preserves line and paragraph continuity that would otherwise be broken. 

\subsection{Gaze Remapping under Screen Magnification}

Screen magnification restricts the visible area and induces systematic shifts in gaze coordinates. We adopt a prior remapping algorithm~\cite{heo2024reading} to project gaze to unmagnified screen coordinates, yielding a consistent representation across magnification settings. This remapping recovers the global layout and preserves line and paragraph continuity that would otherwise be broken. We feed the compensated gaze to the model alongside raw gaze to capture complementary information: raw gaze preserves local oculomotor dynamics tied to the current viewport, whereas the compensated stream stabilizes trajectories in the screen reference frame.

\subsection{Models for Reading Strategy Classification}

Prior work on reading strategy classification has largely relied on hand-crafted features and pre-segmented text. For instance, baselines from ZuCo~\cite{hollenstein2020zuco} required predefined sentence boundaries and computed aggregate gaze features over entire passages. While effective for post-hoc analysis, these approaches are not suited for consequent prediction or interaction. Our method avoids task-specific structure and learns temporal representations that enable fine-grained classification under naturalistic conditions. Statistical analyses of gaze metrics have also been used to compare reading strategies~\cite{gagl2022eye, wallot2011role}, but these provide descriptive insights rather than trainable, generalizable models.

More recently, transformer-based models have been applied to classify reading-related behaviors such as reading, non-reading, and skimming from gaze and additional modalities, including head pose and RGB video~\cite{yang2025reading}. These approaches demonstrate the promise of end-to-end sequence models for behavior classification, but they focus on normally sighted populations and do not consider the challenges introduced by screen magnification. 
\vspace*{-0.3em}
\subsection{Self‑supervised Learning for Gaze and Behavior}
\label{sec:selfsup}

Recent advances in self-supervised learning have led to progress in gaze estimation from eye images~\cite{jindal2023contrastive, wu2022gaze} and in coarse behavior recognition from signals like EOG~\cite{Baray2023EOG, Islam2021SSLReading}. However, little work has been done on learning representations from continuous, high-resolution gaze trajectories. To our knowledge, no prior work has applied self- or semi-supervised learning to frame-level reading behavior classification from dense gaze sequences, particularly in accessibility contexts.

\vspace*{-0.3em}
\subsection{Mouse as Supervisory Signal during Reading}

Mouse behavior has been shown to correlate strongly with gaze during reading, especially in screen-based environments. Cursor movements often reflect user attention and intent, making them a valuable implicit signal. Prior work has aligned mouse and gaze trajectories to analyze attention~\cite{jevremovic2021applying, kirsh2020par} and incorporated cursor features into multimodal pipelines for predicting engagement and reading depth~\cite{cepeda2018mouse, conlen2019capture, huang2012user}. These studies highlight the potential of mouse input for reading behavior analysis, but they primarily focus on alignment or descriptive analysis.

\begin{figure}[t]
  \centering
  \includegraphics[width=\textwidth]{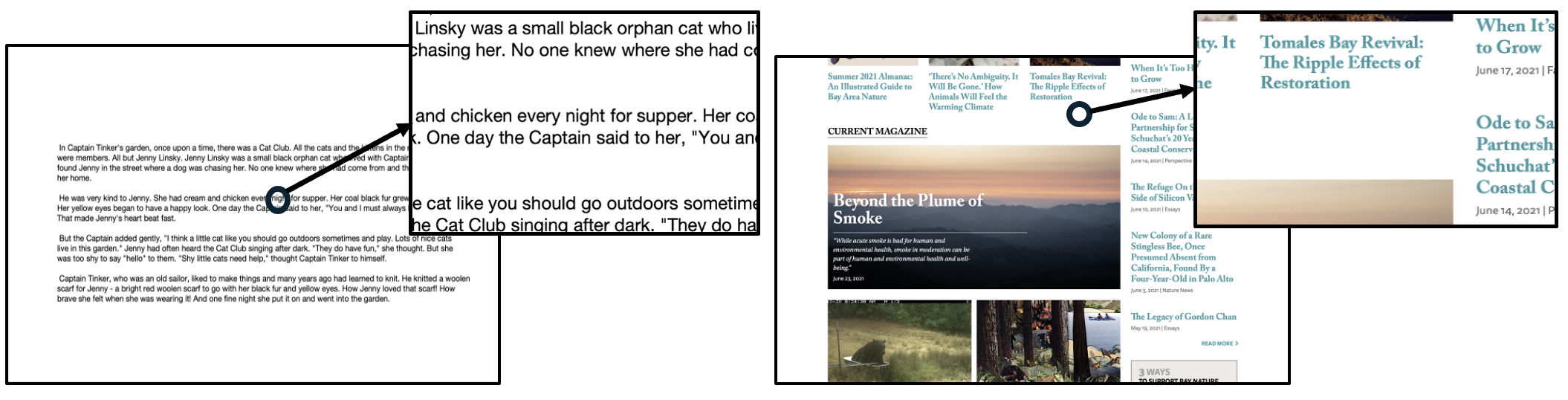}
  \caption{Examples of the two reading tasks (text document and webpage) under the full-lens magnification condition~\cite{tang2023screen}. In this modality, the entire screen is uniformly enlarged, and only a portion of the content is visible at a time, requiring users to continuously scroll to follow the text.}
  \label{fig:dataset}
\end{figure}

\section{Method}
\vspace*{-0.3em}
\subsection{Dataset}

We build on the dataset introduced by Tang et al.\cite{tang2023screen}, which includes synchronized recordings of eye gaze and mouse trajectories from individuals with low vision during screen-based reading. In this work, we focus exclusively on the full-lens magnification condition, where the entire screen is magnified isotropically. Under this setting, users scroll right-to-left (often via left-to-right mouse movements), to keep the line being read within their preferred visual region. Participants read both text documents and webpages, as shown in Fig.~\ref{fig:dataset} recorded with a Tobii Spectrum eye tracker (120Hz) and mouse logs (10Hz). Reading and scanning annotations were produced by trained annotators following a two-pass protocol under expert supervision. 
\vspace*{-0.3em}
\subsection{Model}

\begin{figure}[t]
  \centering
  \includegraphics[width=\textwidth]{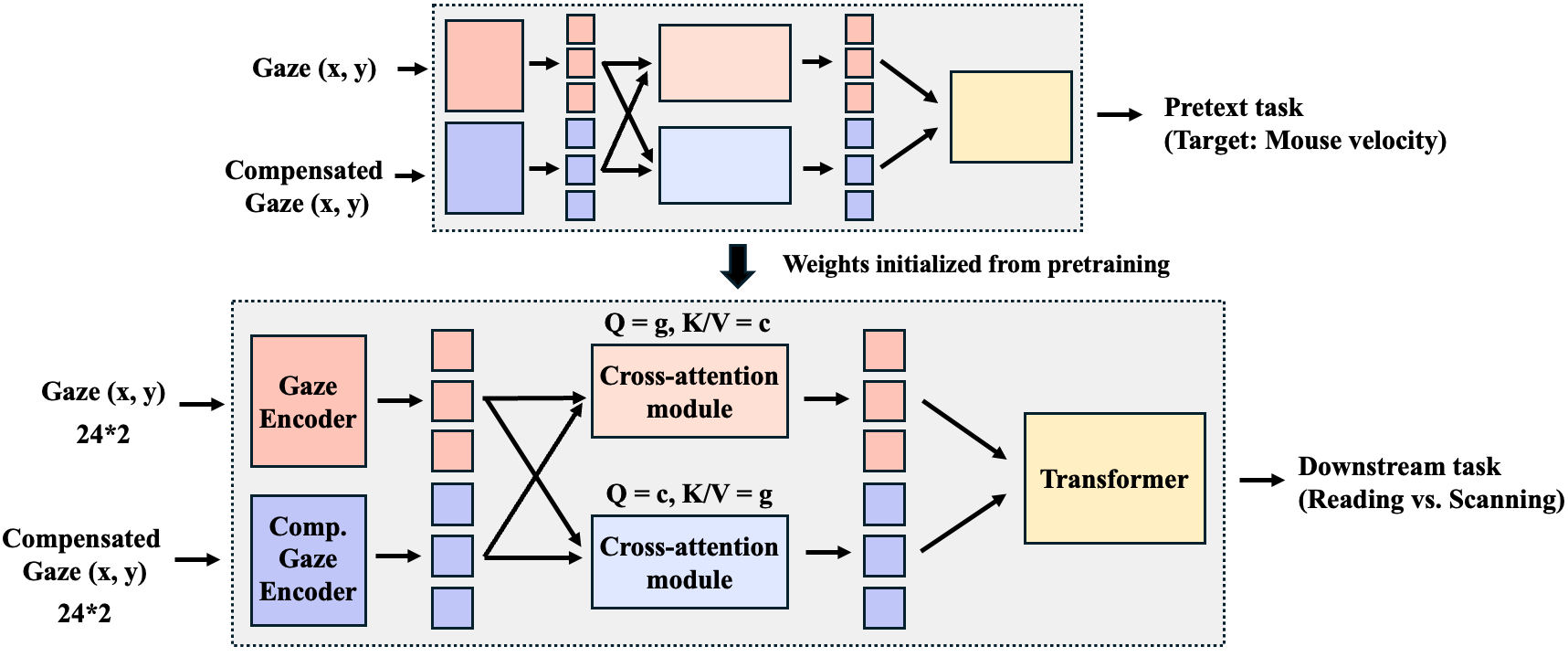}
  \vspace*{-0.5em}
  \caption{Overview of the framework. Top: pretraining (pretext). Bottom: fine-tuning \& inference (downstream). “Comp.” = compensated gaze; g = raw gaze, c = compensated. Both stages share the same backbone: g and c are fused via two cross-attention blocks (Q=g, K/V=c; Q=c, K/V=g) and processed by a Transformer. In pretraining, mouse movements serve only as targets to predict 2-D velocity from gaze; in downstream, the model classifies each sequence as reading or scanning.}
  \label{fig:architecture}
\end{figure}

An overview of the supervised and semi-supervised frameworks is shown in Fig.~\ref{fig:architecture}. We adopted and extended the encoder and Transformer configuration from~\cite{yang2025reading} and ~\cite{heo2025gaze} to suit our task.

\vspace*{-0.5em}

\subsubsection{Supervised Classification}

We first train a reading classifier on labeled gaze data. Input sequences are constructed using overlapping 0.2-second windows (24 steps at 120~Hz), where each window is labeled as reading or scanning based on its final time point. Each window contains two gaze streams: (1) raw gaze in magnified coordinates and (2) compensated gaze remapped to original screen space. Missing values are linearly interpolated, and windows with more than 50\% missing samples are excluded.  

Both gaze streams are independently encoded with a three-layer 1D CNN (kernel size 3, 64 dims) followed by cross-attention to capture complementary information. The fused representation is fed to a three-layer Transformer encoder(64 dims, 4 heads) to model temporal dependencies, and a linear classifier outputs the final label. Weighted cross-entropy is applied to address class imbalance.

\subsubsection{Semi-Supervised Pretraining with Mouse Guidance}

To enrich the gaze encoder with intention-aware features, we add a pretraining stage where the model learns to predict mouse velocity from gaze input. Mouse movements during magnified reading provide a weak supervisory signal that does not require manual behavior labels. The backbone encoder and Transformer are identical across pretraining and fine tuning. Only the task head and loss change: during pretraining, a linear regression head is trained with mean squared error to predict two-dimensional mouse velocity from unlabeled gaze, and during fine-tuning, the head is replaced with a classifier trained with cross entropy on intent labels. 

\section{Experiments and Results}
To ensure subject-independent evaluation, we adopt a leave-one-subject-out cross-validation scheme, holding out each participant in turn for testing while training on the remaining subjects. All models were trained with the Adam optimizer (lr = 3e-4, weight decay = 0.01).
\subsection{Supervised Ablation on Gaze and Mouse Inputs}

\textbf{Supervised Learning with Gaze Input} \hspace{0.5em} 
Table~\ref{tab:multi_modal_input_results} compares three input configurations. Raw gaze and compensated gaze provide complementary views: raw captures fine-grained oculomotor dynamics within the magnified viewport, while compensated gaze restores global spatial consistency. Their fusion achieves the best gaze-only performance (80.02 F1), substantially higher than raw-only (75.06) or compensated-only (67.22). We select the eye (left or right) with the lowest NaN ratio in the first 10\% of the session as a lightweight calibration heuristic. The random-label baseline (40.91 F1) confirms non-trivial of the task.

\textbf{Supervised Learning with Mouse Input} \hspace{0.5em} 
We first verify that mouse trajectories themselves provide meaningful behavioral signals. Mouse-only input is weaker than gaze (52.85 F1), yet combining it with gaze and compensated gaze improves performance to 83.64 F1. This shows mouse trajectories carry useful behavioral cues. Scanning F1 in particular improves markedly (68.78 $\rightarrow$ 76.10), indicating that mouse trajectories provide strong cues for exploratory behavior such as line skipping or reorientation. Since our target is hands-free interaction, however, mouse input is not available at inference. This motivates a semi-supervised approach in which mouse data serves only as auxiliary supervision during pretraining. 

\subsection{Semi-Supervised Gaze Representation Learning under Mouse Supervision}

\begin{table}[t]
\centering
\caption{F1 scores on the text dataset for different input configurations. "Comp." denotes compensated gaze. "Random" uses permuted labels with gaze-only input while preserving the original class distribution, serving as a sanity check. \textbf{Bold} indicates the best performance among gaze-only inputs.}
\resizebox{0.95\linewidth}{!}{
\begin{tabular}{l|cccc|cc}
\toprule
\textbf{Metric} 
& \multicolumn{4}{c|}{\textit{Gaze-Only Inputs}} 
& \multicolumn{2}{c}{\textit{With Mouse Inputs}} \\
& Gaze + Comp. & Gaze & Comp. & Random 
& Mouse & Mouse + Gaze + Comp. \\
\midrule
Overall F1     & \textbf{80.02} & 75.06 & 67.22 & 40.91 & 52.85 & 83.64 \\
Reading F1     & \textbf{91.27} & 89.31 & 87.69 & 56.04 & 70.29 & 91.17\\
Scanning F1    & \textbf{68.78} & 60.81 & 46.75 & 25.79 & 35.41 & 76.10 \\
\bottomrule
\end{tabular}}
\label{tab:multi_modal_input_results}
\end{table}

\begin{table}[t]
\centering
\caption{F1 scores of supervised and semi-supervised under two strategies (partial vs. full). 'Partial' updates only the last three Transformer layers, while "Full" fine-tunes the entire model. \textbf{Bold} indicates the best performance.}
\resizebox{0.95\linewidth}{!}{
\begin{tabular}{llcccccc}
\toprule
\textbf{Input Type} & \multicolumn{3}{c}{\textbf{Text}} & \multicolumn{3}{c}{\textbf{Webpage}} \\
\cmidrule(lr){2-4} \cmidrule(lr){5-7}
&  Overall & Reading & Scanning & Overall & Reading & Scanning \\
\midrule
Semi-supervised (Partial) & 81.93 & 91.56 & 72.29 & 64.51 & 62.59 & 66.42 \\ % & 52.22 & 49.43 & 55.00 \\
Semi-supervised (Full)  & \textbf{85.97} & \textbf{93.13} & \textbf{78.80} & \textbf{70.01} & \textbf{68.39} & \textbf{71.62} \\ % & 55.02 & 51.23 & 58.81 \\
\midrule
Supervised & 80.02 & 91.27 & 68.78 & 62.49 & 60.27 & 64.59 \\
\bottomrule
\end{tabular}}
\label{tab:semi_results_combined}
\end{table}

We pretrain with a mouse-guided gaze reconstruction objective and fine-tune on labeled data using two strategies: \emph{partial}, which updates only the top three Transformer layers, and \emph{full}, which updates all parameters. Both variants outperform the supervised baseline, demonstrating the utility of mouse signals as weak supervision. Even partial adaptation exceeds supervised training, underscoring the quality of the learned representations.

On text documents, full fine-tuning improves overall F1 by 6.0\% (80.02 $\rightarrow$ 85.97). On the more challenging webpage setting, gains are larger: 7.5\% (62.49 $\rightarrow$ 70.01). These results show that mouse-guided pretraining yields transferable gaze encoders, with benefits amplified under complex, variable environments.

\section{Conclusion}

We introduced a semi-supervised learning framework that uses mouse behavior to guide the learning of gaze representations for reading behavior classification under screen magnification. By pretraining on mouse-guided prediction and fine-tuning on labeled gaze, our method substantially improves over fully supervised baselines. The joint modeling of raw and compensated gaze further enhances robustness, capturing both fine-grained oculomotor signals and consistent global patterns. Experiments on both text and webpage tasks demonstrate that mouse-guided pretraining is especially beneficial in noisy, complex environments. Importantly, inference remains gaze-only, making the approach practical for real-world, hands-free accessibility applications.

For future work, we plan to translate our intent classifier into a real-time, fully automatic scroll controller that interprets reading and scanning probabilities to generate smooth, safe viewport movements, leveraging hysteresis and uncertainty-aware mechanisms to enable robust hands-free accessibility on standard devices.

\section*{Acknowledgement}
Special thanks to Tejas Polu, Aswhin Nagarajan, Suhas Oruganti, Shalini Raval, and Arav Adhikari for their assistance with data annotation and code debugging.

\bibliographystyle{plain}
\bibliography{ref.bib}

\end{document}